# The Power of Nondeterminism in Self-Assembly*

Nathaniel Bryans†    Ehsan Chiniforooshan†    David Doty‡    Lila Kari†

Shinnosuke Seki†

## Abstract

We investigate the role of nondeterminism in Winfree's abstract Tile Assembly Model (aTAM), which was conceived to model artificial molecular self-assembling systems constructed from DNA. Designing tile systems that assemble shapes, due to the algorithmic richness of the aTAM, is a form of sophisticated "molecular programming". Of particular practical importance is to find tile systems that minimize resources such as the number of distinct tile types, each of which corresponds to a set of DNA strands that must be custom-synthesized in actual molecular implementations of the aTAM. We seek to identify to what extent the use of nondeterminism in tile systems affects the resources required by such molecular shape-building algorithms.

By nondeterminism we do not mean a magical ability such as that possessed by a nondeterministic algorithm to search an exponential-size space in polynomial time. Rather, we study realistically implementable systems that retain a different sense of determinism in that they are guaranteed to produce a unique shape but are nondeterministic in that they do not guarantee which tile types will be placed where within the shape. A sensible analogy is a nondeterministic algorithm that outputs the same value on all computation paths for a given input. Such an algorithm can always be replaced by an equivalent deterministic algorithm with the same running time, memory usage, and program length. It is then intuitively reasonable to conjecture that a similar equivalence should hold between deterministic tile systems and those nondeterministic tile systems that always "output" the same shape.

This intuition is wrong. We first show a "molecular computability theoretic" result: there is an infinite shape $S$ that is uniquely assembled by a tile system but not by any deterministic tile system. We then show an analogous phenomenon – using a different technique – in the finitary "molecular complexity theoretic" case: there is a finite shape $S$ that is uniquely assembled by a tile system with $c$ tile types, but every deterministic tile system that uniquely assembles $S$ has more than $c$ tile types. In fact we extend the technique to derive a stronger (classical complexity theoretic) result, showing that the problem of finding the minimum number of tile types that uniquely assemble a given finite shape is $\Sigma_2^\mathsf{P}$-complete. In contrast, the problem of finding the minimum number of *deterministic* tile types that uniquely assemble a shape was shown to be NP-complete by Adleman, Cheng, Goel, Huang, Kempe, Moisset de Espanés, and Rothemund (*Combinatorial Optimization Problems in Self-Assembly*, STOC 2002).

The conclusion is that nondeterminism confers extra power to assemble a shape from a small tile system, but unless the polynomial hierarchy collapses, it is computationally more difficult to exploit this power by finding the size of the smallest tile system, compared to finding the size of the smallest deterministic tile system.

*This research was supported by NSERC Discovery Grant R2824A01 and the Canada Research Chair Award in Biocomputing to Lila Kari, by the NSERC Undergraduate Student Research Awards (USRA) grant to Nathaniel Bryans, and by the NSF Computing Innovation Fellowship grant to David Doty.

†University of Western Ontario, Dept. of Computer Science, London, Ontario, Canada, N6A 5B7, `{nbryans,ehsan,lila,sseki}@csd.uwo.ca`.

‡California Institute of Technology, Dept. of Computer Science, Pasadena, CA 91125, USA, `ddoty@caltech.edu`

# 1 Introduction

Tile self-assembly is an algorithmically rich model of "programmable crystal growth". It is possible to design molecules (square-like "tiles") with specific binding sites so that, even subject to the chaotic nature of molecules floating randomly in a well-mixed chemical soup, they are guaranteed to bind so as to deterministically form a single target shape. This is despite the number of different types of tiles possibly being much smaller than the size of the shape and therefore having only "local information" to guide their attachment. The ability to control nanoscale structures and machines to atomic-level precision will rely crucially on sophisticated self-assembling systems that automatically control their own behavior where no top-down externally controlled device could fit.

A practical implementation of self-assembling molecular tiles was proved experimentally feasible in 1982 by Seeman [38] using DNA complexes formed from artificially synthesized strands. Experimental advances have delivered increasingly reliable assembly of algorithmic DNA tiles with error rates of 10% per tile in 2004 [36], 1.4% per tile in 2007 [17], and 0.13% per tile in 2009 [8]. Erik Winfree [44] introduced the abstract Tile Assembly Model (aTAM) – based on a constructive version of Wang tiling [42, 43] – as a simplified mathematical model of self-assembling DNA tiles. Winfree demonstrated the computational universality of the aTAM by showing how to simulate an arbitrary cellular automaton with a tile assembly system. Building on these connections to computability, Rothemund and Winfree [35] investigated a self-assembly resource bound known as *tile complexity*, the minimum number of tile types needed to assemble a shape. They showed that for most $n$, the problem of assembling an $n \times n$ square has tile complexity $\Omega\left(\frac{\log n}{\log \log n}\right)$, and Adleman, Cheng, Goel, and Huang [4] exhibited a construction showing that this lower bound is asymptotically tight. Under natural generalizations of the model [2, 6, 9–13, 19, 20, 29, 39, 41], tile complexity can be reduced for tasks such as square-building and assembly of more general shapes.

There are different interpretations of "nondeterminism" in the aTAM. We say a tile system is *directed (a.k.a. deterministic)* if it is guaranteed to form one unique final assembly, where an assembly is defined not only by which positions are eventually occupied by a tile, but also by which tile type is placed at each position. We say a tile system *strictly (a.k.a. uniquely) self-assembles* a shape if all of its final assemblies are guaranteed to have that shape. A natural analogy may be made between a non-directed tile system that strictly self-assembles some shape and a nondeterministic Turing machine $N$ that always produces the same output on a given input, regardless of the nondeterministic choices made during computation. There is always a deterministic Turing machine $M$ computing the same function as $N$ and using no more "resources", according to any common resource bound such as time complexity, space complexity, or program length. Therefore we regard such a restricted class of nondeterministic Turing machines as no more "powerful" than deterministic Turing machines.

Based on this analogy, it might seem that strict self-assembly, while allowing one form of nondeterminism (which tile goes where), so strongly requires another form of determinism (which positions have a tile) that extra power cannot be gained by allowing the tile systems to be non-directed. More precisely, it is natural to conjecture that every infinite shape that is strictly self-assembled by some tile system is also strictly self-assembled by some directed tile system. In the finitary case, *every* finite shape is assembled by a directed tile system (possibly using as many tile types as there are points in the shape), so to make the idea non-trivial we might conjecture that the tile complexity of a finite shape is independent of whether we consider all tile systems or only those that are directed. Such conjectures are appealing because the algorithmic design and verification of tile systems [39] as well as lower bounds and impossibility proofs [6, 15, 30] often rely on reasoning about directed tile systems, which are "better behaved" in many senses than arbitrary tile systems, even those that strictly self-assemble a shape. It would be helpful to begin such arguments with the phrase, "Assume without loss of generality that the tile system is directed."

However, these conjectures are false. We show that there is an infinite shape $S$ that is strictly self-assembled by a tile system but not by any directed tile system. Therefore, in a "molecular computability theoretic" sense, nondeterminism allows certain shapes to be algorithmically self-assembled that are totally "unassemblable" (to borrow Adleman's tongue-twisting analog of "uncomputable" [3]) under the constraint of determinism. We then show an analogous phenomenon in the finitary case: there is a finite shape $S$ that is strictly self-assembled by a tile system with $c$ tile types, but every directed tile system that strictly self-assembles $S$ has more than $c$ tile types. In fact to strictly self-assemble the shape in a directed tile system requires more than $\approx \frac{3}{2}c$ tile types. It is open to improve the constant $\frac{3}{2}$ to be larger or to prove a super-linear gap between the complexity measures; the issue is discussed in more detail in Section 5. This establishes a "molecular complexity theoretic" analog of the first result. We then derive a stronger result, showing that the problem of finding the minimum number of tile types that strictly self-assemble a given finite shape is complete for the complexity class $\Sigma_2^P = NP^{NP}$. In contrast, the problem of finding the minimum number of *directed* tile types that strictly self-assemble a shape was shown to be NP-complete by Adleman, Cheng, Goel, Huang, Kempe, Moisset de Espanés, and Rothemund [5].

Based on these results, we conclude that nondeterminism confers extra power to assemble a shape from a small tile system, but unless the polynomial hierarchy collapses, it is computationally more difficult[1] to exploit this power by finding the size of the smallest tile system, compared to finding the size of the smallest directed tile system.

One might argue that this difference between nondeterministic (but "output-deterministic") Turing machines and non-directed (but strict) tile systems is not surprising, since there is a "monotone" aspect to tile assembly in the sense that space used to place a tile cannot be reused, whereas a tape cell used to store information by a Turing machine can be reused to store different information later. It is sometimes said that the difference between space and time is that "you can reuse space but you cannot reuse time." However, the "computation" carried out by tile systems does not distinguish well between space and time. For instance, the standard simulation of a Turing machine by a tile system (see [35]) assembles a structure encoding the entire space-time configuration history of the Turing machine. Even with negative glue strengths that are able to force detachments to occur, the volume requirements of such a simulation must be proportional to $t \cdot s$ for a Turing machine using time $t$ and space $s$ [14], essentially forcing the solution to contain multiple assemblies that collectively encode the entire computation history. Tile systems therefore cannot reuse space (tiles), which is the fundamental effect of their monotonicity on their computational abilities. From this perspective, a Turing machine cannot reuse time any better than any other computational system (barring the use of closed timelike curves [1]), yet in contrast to tile systems, a nondeterministic but "output-deterministic" Turing machine remains no more powerful, even in the sense of time complexity, than a deterministic Turing machine.

---

[1] "More difficult" in the sense of nondeterministic time complexity, although it is conceivable that both problems have the same deterministic time complexity.

## 2 Abstract Tile Assembly Model

This section gives a terse definition of the abstract Tile Assembly Model (aTAM, [44]). This is not a tutorial; for readers unfamiliar with the aTAM, [35] gives an excellent introduction to the model.

Fix an alphabet $\Sigma$. $\Sigma^*$ is the set of finite strings over $\Sigma$. Given a discrete object $O$, $\langle O \rangle$ denotes a standard encoding of $O$ as an element of $\Sigma^*$. $\mathbb{Z}$, $\mathbb{Z}^+$, and $\mathbb{N}$ denote the set of integers, positive integers, and nonnegative integers, respectively. For a set $A$, $\mathcal{P}(A)$ denotes the power set of $A$. Given $A \subseteq \mathbb{Z}^2$, the *full grid graph* of $A$ is the undirected graph $G_A^{\mathrm{f}} = (V, E)$, where $V = A$, and for all $u, v \in V$, $\{u, v\} \in E \iff \|u - v\|_2 = 1$; i.e., iff $u$ and $v$ are adjacent on the integer Cartesian plane. A *shape* is a set $S \subseteq \mathbb{Z}^2$ such that $G_S^{\mathrm{f}}$ is connected. A shape $\Upsilon$ is a *tree* if $G_\Upsilon^{\mathrm{f}}$ is acyclic.

A *tile type* is a tuple $t \in (\Sigma^* \times \mathbb{N})^4$; i.e., a unit square with four sides listed in some standardized order, each side having a *glue* $g \in \Sigma^* \times \mathbb{N}$ consisting of a finite string *label* and nonnegative integer *strength*. We assume a finite set $T$ of tile types, but an infinite number of copies of each tile type, each copy referred to as a *tile*. An *assembly* is a nonempty connected arrangement of tiles on the integer lattice $\mathbb{Z}^2$, i.e., a partial function $\alpha : \mathbb{Z}^2 \dashrightarrow T$ such that $G_{\mathrm{dom}\,\alpha}^{\mathrm{f}}$ is connected and $\mathrm{dom}\,\alpha \neq \varnothing$. The *shape* $S_\alpha \subseteq \mathbb{Z}^2$ of $\alpha$ is $\mathrm{dom}\,\alpha$. Two adjacent tiles in an assembly *interact* if the glues on their abutting sides are equal (in both label and strength) and have positive strength. Each assembly $\alpha$ induces a *binding graph* $G_\alpha^{\mathrm{b}}$, a grid graph whose vertices are positions occupied by tiles, with an edge between two vertices if the tiles at those vertices interact.[2] Given $\tau \in \mathbb{Z}^+$, $\alpha$ is $\tau$-*stable* if every cut of $G_\alpha^{\mathrm{b}}$ has weight at least $\tau$, where the weight of an edge is the strength of the glue it represents. That is, $\alpha$ is $\tau$-stable if at least energy $\tau$ is required to separate $\alpha$ into two parts. When $\tau$ is clear from context, we say $\alpha$ is *stable*. Given two assemblies $\alpha, \beta : \mathbb{Z}^2 \dashrightarrow T$, we say $\alpha$ is a *subassembly* of $\beta$, and we write $\alpha \sqsubseteq \beta$, if $S_\alpha \subseteq S_\beta$ and, for all points $p \in S_\alpha$, $\alpha(p) = \beta(p)$.

A *tile assembly system* (TAS) is a triple $\mathcal{T} = (T, \sigma, \tau)$, where $T$ is a finite set of tile types, $\sigma : \mathbb{Z}^2 \dashrightarrow T$ is the finite, $\tau$-stable *seed assembly*, and $\tau \in \mathbb{Z}^+$ is the *temperature*. Given two $\tau$-stable assemblies $\alpha, \beta : \mathbb{Z}^2 \dashrightarrow T$, we write $\alpha \to_1^{\mathcal{T}} \beta$ if $\alpha \sqsubseteq \beta$ and $|S_\beta \setminus S_\alpha| = 1$. In this case we say $\alpha$ $\mathcal{T}$-*produces* $\beta$ *in one step*.[3] If $\alpha \to_1^{\mathcal{T}} \beta$, $S_\beta \setminus S_\alpha = \{p\}$, and $t = \beta(p)$, we write $\beta = \alpha + (p \mapsto t)$. The $\mathcal{T}$-*frontier* of $\alpha$ is the set $\partial^{\mathcal{T}}\alpha = \bigcup_{\alpha \to_1^{\mathcal{T}} \beta} S_\beta \setminus S_\alpha$, the set of empty locations at which a tile could stably attach to $\alpha$.

A sequence of $k \in \mathbb{Z}^+ \cup \{\infty\}$ assemblies $\alpha_0, \alpha_1, \ldots$ is a $\mathcal{T}$-*assembly sequence* if, for all $1 \leq i < k$, $\alpha_{i-1} \to_1^{\mathcal{T}} \alpha_i$. We write $\alpha \to^{\mathcal{T}} \beta$, and we say $\alpha$ $\mathcal{T}$-*produces* $\beta$ (in 0 or more steps) if there is a $\mathcal{T}$-assembly sequence $\alpha_0, \alpha_1, \ldots$ of length $k = |S_\beta \setminus S_\alpha| + 1$ such that 1) $\alpha = \alpha_0$, 2) $S_\beta = \bigcup_{0 \leq i < k} S_{\alpha_i}$, and 3) for all $0 \leq i < k$, $\alpha_i \sqsubseteq \beta$. If $k$ is finite then it is routine to verify that $\beta = \alpha_{k-1}$.[4] We say $\alpha$ is $\mathcal{T}$-*producible* if $\sigma \to^{\mathcal{T}} \alpha$, and we write $\mathcal{A}[\mathcal{T}]$ to denote the set of $\mathcal{T}$-producible assemblies. The relation $\to^{\mathcal{T}}$ is a partial order on $\mathcal{A}[\mathcal{T}]$ [23, 34]. A $\mathcal{T}$-assembly sequence $\alpha_0, \alpha_1, \ldots$ is *fair* if, for all $i$ and all $p \in \partial^{\mathcal{T}}\alpha_i$, there exists $j$ such that $\alpha_j(p)$ is defined; i.e., no frontier location is "starved".

An assembly $\alpha$ is $\mathcal{T}$-*terminal* if $\alpha$ is $\tau$-stable and $\partial^{\mathcal{T}}\alpha = \varnothing$. We write $\mathcal{A}_\square[\mathcal{T}] \subseteq \mathcal{A}[\mathcal{T}]$ to denote the set of $\mathcal{T}$-producible, $\mathcal{T}$-terminal assemblies. A TAS $\mathcal{T}$ is *directed (a.k.a., deterministic,*

---

[2] For $G_{S_\alpha}^{\mathrm{f}} = (V_{S_\alpha}, E_{S_\alpha})$ and $G_\alpha^{\mathrm{b}} = (V_\alpha, E_\alpha)$, $G_\alpha^{\mathrm{b}}$ is a spanning subgraph of $G_{S_\alpha}^{\mathrm{f}}$: $V_\alpha = V_{S_\alpha}$ and $E_\alpha \subseteq E_{S_\alpha}$.

[3] Intuitively $\alpha \to_1^{\mathcal{T}} \beta$ means that $\alpha$ can grow into $\beta$ by the addition of a single tile; the fact that we require both $\alpha$ and $\beta$ to be $\tau$-stable implies in particular that the new tile is able to bind to $\alpha$ with strength at least $\tau$. It is easy to check that had we instead required only $\alpha$ to be $\tau$-stable, and required that the cut of $\beta$ separating $\alpha$ from the new tile has strength at least $\tau$, then this implies that $\beta$ is also $\tau$-stable.

[4] If we had defined the relation $\to^{\mathcal{T}}$ based on only finite assembly sequences, then $\to^{\mathcal{T}}$ would be simply the reflexive, transitive closure $(\to_1^{\mathcal{T}})^*$ of $\to_1^{\mathcal{T}}$. But this would mean that no infinite assembly could be produced from a finite assembly, even though there is a well-defined, unique "limit assembly" of every infinite assembly sequence.

*confluent*) if the poset $(\mathcal{A}[\mathcal{T}], \to^{\mathcal{T}})$ is directed; i.e., if for each $\alpha, \beta \in \mathcal{A}[\mathcal{T}]$, there exists $\gamma \in \mathcal{A}[\mathcal{T}]$ such that $\alpha \to^{\mathcal{T}} \gamma$ and $\beta \to^{\mathcal{T}} \gamma$.[5] We say that a TAS $\mathcal{T}$ *strictly (a.k.a. uniquely) self-assembles* a shape $S \subseteq \mathbb{Z}^2$ if, for all $\alpha \in \mathcal{A}_\square[\mathcal{T}]$, $S_\alpha = S$; i.e., if every terminal assembly produced by $\mathcal{T}$ has shape $S$. If $\mathcal{T}$ strictly self-assembles some shape $S$, we say that $\mathcal{T}$ is *strict*. Note that the implication "$\mathcal{T}$ is directed $\implies$ $\mathcal{T}$ is strict" holds, but the converse does not hold.

In this paper we will always use *singly-seeded temperature-2* TAS's, those with $|S_\sigma| = 1$ and $\tau = 2$; hence we will use the term *seed tile* for $\sigma$ as well, and for the remainder of this paper we use the term TAS to mean singly-seeded temperature-2 TAS. When $\mathcal{T}$ is clear from context, we may omit $\mathcal{T}$ from the notation above and instead write $\to_1$, $\to$, $\partial \alpha$, *frontier*, *assembly sequence*, *produces*, *producible*, and *terminal*. Since the behavior of a TAS $\mathcal{T} = (T, \sigma, 2)$ is unchanged if every glue with strength greater than 2 is changed to have strength exactly 2, we assume henceforth that all glue strengths are 0, 1, or 2, and use the terms *null glue*, *single glue*, and *double glue*, respectively, to refer to these three cases.[6] We also assume without loss of generality that every single glue or double glue occurring in some tile type in some direction also occurs in some tile type in the opposite direction, i.e., there are no "effectively null" single or double glues.[7]

## 3 Assembly of Infinite Shapes

In this section we study the power of nondeterminism in assembling infinite shapes. The following theorem is the main result of Section 3.

**Theorem 3.1.** *There is a shape $S \subset \mathbb{Z}^2$ such that some TAS strictly self-assembles $S$, but no directed TAS strictly self-assembles $S$.*

*Proof.* Let $L \subset \mathbb{N}$ be a language that is computably enumerable but not decidable, and let $M$ be a Turing machine such that $L = L(M)$. Let $S$ be the shape that is strictly self-assembled by the TAS described below, when $M$ is encoded into the TAS as described.

A portion of the shape $S$ is shown in Figure 1. The TAS that strictly self-assembles $S$ is based on the main construction of Lathrop, Lutz, Patitz, and Summers [22]. In that paper, the authors show that for each Turing machine $M$, an encoding of the language $L(M) \subseteq \mathbb{N}$ accepted by $M$ "weakly self-assembles" on the $x$-axis. More precisely, for a "reasonably simple" function $f : \mathbb{N} \to \mathbb{N}$, a special tile type is placed at position $(f(n), 0)$ if and only if $n \in L(M)$. The $n^{\text{th}}$ "ray" in Figure 1 begins growth just before $(f(n), 0)$, and grows independently of the other rays, controlling an adjacent simulation of $M(n)$ in parallel with all the other rays. The slope of each ray is just a bit smaller than the previous, with the slope approaching 2 as $n \to \infty$. The simulation executes one transition of $M$ on input $n$ every $\approx 2^n$ rows of the ray. Since $M$ can use no more than $k$ tape cells after $k$ transitions, this slowed simulation ensures that each ray has enough space to allow a potentially unbounded simulation of $M$ on each $n$, without "crashing" into the next adjacent ray, even in the worst case that $M$ moves its tape head right on every transition.

What is needed from this construction for our purpose is:

---
[5]The following two convenient characterizations of "directed" are routine to verify. $\mathcal{T}$ is directed if and only if $|\mathcal{A}_\square[\mathcal{T}]| = 1$. $\mathcal{T}$ is *not* directed if and only if there exist $\alpha, \beta \in \mathcal{A}[\mathcal{T}]$ and $p \in S_\alpha \cap S_\beta$ such that $\alpha(p) \neq \beta(p)$.

[6]We use *null bond*, *single bond*, and *double bond* similarly to refer to the *interaction* of two tiles.

[7]Thus the existence of a tile with a double glue facing empty space implies that the empty space is part of the frontier. Many of our arguments use the contrapositive that if a shape $S$ is strictly self-assembled by a tile system and a side of a tile faces a point $p \notin S$, then the tile cannot have a double glue on that side.

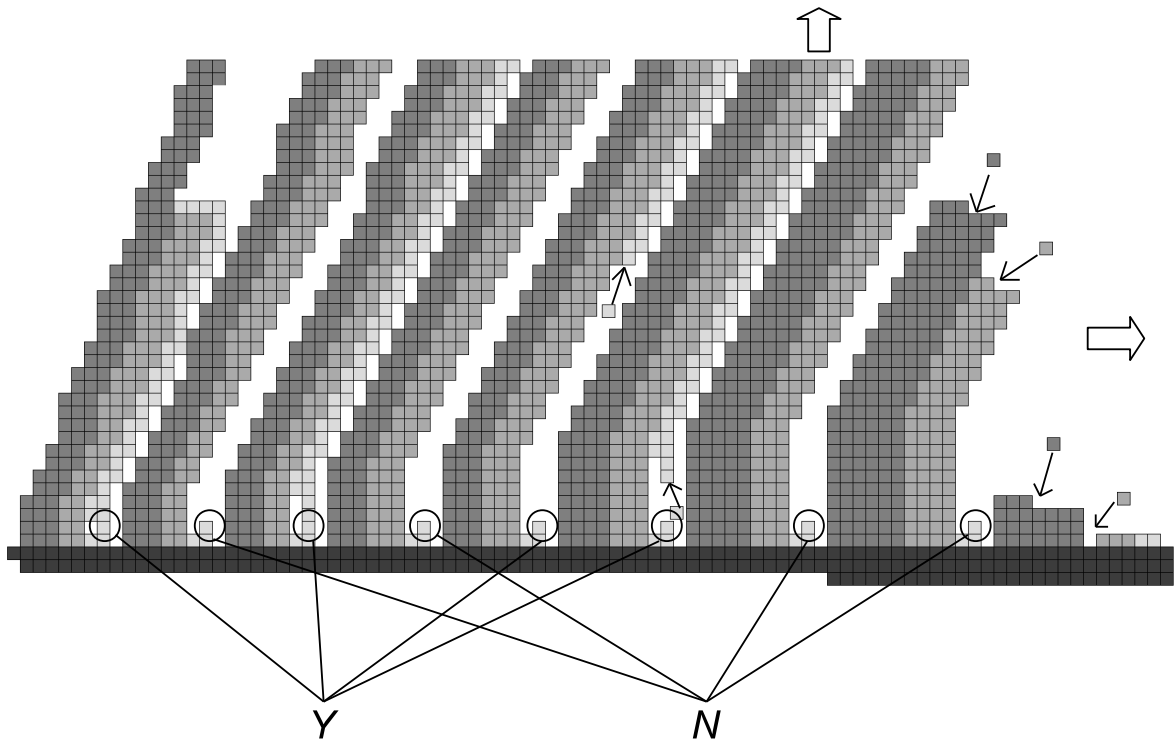

Figure 1: A portion of an infinite shape $S$ that strictly self-assembles, but not by any directed TAS. The $n^{\text{th}}$ ray simulates a Turing machine $M$ on input $n$, and a vertical line is present under that ray if and only if $M$ accepts $n$. $Y$ and $N$ are points at these positions representing "yes" and "no" instances of $L(M)$, respectively, and elements of $Y$ are the points where nondeterminism is forced to occur in any TAS that strictly self-assembles $S$.

1. $f$ is computable.[8]

2. The simulation of $M(n)$, carried out adjacent to the $n^{\text{th}}$ ray, sends a "signal" crawling down the right side of the simulation if and only if $M$ accepts $n$, placing a special tile just above the "planter" (the group of tiles growing below each of the rays).

We modify the signal so that, rather than growing all the way to the planter, for input $n$, the signal grows to distance $n$ north of the planter and then grows a width-1 vertical line $n$ positions down to the planter, using the same tile type with equal north and south double glues to "crash" into the planter. To ensure that the downward-growing vertical lines do not obstruct the operation of the planter, the planter is modified so that it is guaranteed to grow horizontally a sufficient number of tiles before laying out the input for the $M$, so as to guarantee that there is something present for a "controlled crash." The space for the downward-growing line of length $n$ to drop after the input is accepted is created by having the Turing machine simulations begin not immediately above the planter, but at height $n$ on input $n$. This is why the $n^{\text{th}}$ ray grows straight up for $n$ rows before beginning its sloped growth. Under *every* simulation, a "notch" tile is placed above the planter using a double glue, which is horizontally lined up with where the vertical line will grow if $M$ accepts. The actions of the ray, planter and Turing machine simulation are otherwise similar to the mechanisms used in [22]. We note that this particular TAS is not directed since the "notch" tiles compete nondeterministically with the vertical line tiles at positions where $M$ accepts.

---

[8] [22] defines the roughly quadratic function $f(n) = \binom{n+1}{2} + (n+1)\lfloor \log n \rfloor + 6n - 2^{1+\lfloor \log n \rfloor} + 2$. Our version of this function will grow just a bit faster, to make room for a vertical line to form between two adjacent rays without "touching" the rest of the shape except at the endpoints of the line, but retains computability.

It remains to show that no directed TAS strictly self-assembles $S$. Intuitively, we show that at points of the form $(f(n), 0)$, any directed TAS must place tiles that "know" whether there will eventually be a vertical line above the point, implying the ability to decide $L$ since vertical lines appear above exactly those positions $(f(n), 0)$ such that $n \in L$. Assume for the sake of contradiction that there is a directed TAS $\mathcal{T} = (T, \sigma, 2)$ that strictly self-assembles $S$, and let $\alpha \in \mathcal{A}_\square[\mathcal{T}]$ be its unique producible, terminal assembly. Since the heights of the vertical "bases" of each ray below the sloped portion are strictly increasing, there is some $n_0 \in \mathbb{N}$ such that, for all $n > n_0$, the distance from $(f(n), 1)$ to the ray above it is at least $|T| + 1$. Let $Y = \{\ (f(n), 0) \mid n \in L \text{ and } n > n_0\ \}$ be the bottommost points of the vertical lines adjacent to rays corresponding to (sufficiently large) "yes" instances of $L$, and let $N = \{\ (f(n), 0) \mid n \notin L \text{ and } n > n_0\ \}$ represent the positions of the "notches" corresponding to (sufficiently large) "no" instances. $Y$ and $N$ are shown in Figure 1. Let $T_Y = \alpha(Y)$ and $T_N = \alpha(N)$ be the set of tile types that appear at "yes" and "no" instance points, respectively. Since $S$ has empty space immediately north of positions in $N$, no tile type in $T_N$ has a north double glue.

We claim that $T_Y \cap T_N = \varnothing$. For the sake of contradiction, suppose otherwise, let $t \in T_Y \cap T_N$, and let $p \in Y$ be a point where $\alpha(p) = t$. Since $t \in T_N$, $t$ has no north double glue, so the vertical line above $p$ must grow downward using north and south double glues. Let $q = p + (0, 1)$ be the point just above $p$. By our choice of $n_0$, the vertical line must repeat a tile type before reaching the point $q$, so all tile types in the repetition period have a north and a south double glue, including the tile type $t' = \alpha(q)$. Let $t''$ be the tile type appearing beneath $t'$ after the previous occurrence of $t'$ in the vertical line. Since $t''$ has a north double glue, $t'' \notin T_N$, so $t'' \neq t$. Because $t$ binds to the rest of $\alpha$ only through its south double glue, there can be no precedence relationship enforcing that $p$ must contain a tile before $q$ (or any other point) receives a tile. In other words, there exists a producible assembly $\beta \in \mathcal{A}[\mathcal{T}]$ such that $\beta(q) = t'$ and $\beta(p)$ is undefined. This implies that $t''$ can bind to $\beta$ at position $p$ to create $\beta' = \beta + (p \mapsto t'')$, contradicting the directedness of $\mathcal{T}$ since $\beta', \alpha \in \mathcal{A}[\mathcal{T}]$ but $\beta'(p) = t'' \neq t = \alpha(p)$. This verifies the claim that $T_Y \cap T_N = \varnothing$.

For all $n \in \mathbb{N}$, let $p_n = (f(n), 0)$. Since $T_Y \cap T_N = \varnothing$, for all $n > n_0$, $n \in L \iff p_n \in Y \iff \alpha(p_n) \in T_Y$, and $n \notin L \iff p_n \in N \iff \alpha(p_n) \in T_N$. Using this fact, we describe an algorithm to decide $L$, contradicting its undecidability and completing the proof. On input $n \in \mathbb{N}$, if $n \leq n_0$, use a constant lookup table to decide $n$. Otherwise, compute $p_n = (f(n), 0)$. Simulate the assembly of $\mathcal{T}$ with a fair assembly sequence, maintaining a first-in, first-out queue of frontier locations to enforce fairness, until a tile is placed at position $p_n$. Since this assembly sequence is fair, the simulation will eventually place a tile type $\alpha(p_n)$ at $p_n$, and $\alpha(p_n)$'s membership in $T_Y$ or $T_N$ will indicate whether to accept or reject $n$. □

We have implemented the tile assembly system that strictly self-assembles $S$:
http://www.dna.caltech.edu/~ddoty/pnsa/
It can be simulated using Matthew Patitz's ISU TAS simulator [33] available here:
http://www.cs.iastate.edu/~lnsa/software.html

The purpose of the implementation is not to quantitatively analyze the construction, since we make no quantitative claims about either the shape being assembled nor the TAS that strictly self-assembles the shape. Furthermore, the bulk of the intellectual effort in the proof of Theorem 3.1 is proving the negative result that no directed TAS strictly self-assembles the shape, which is something that cannot be established through a simulation. We provide the simulation primarily to help the interested reader understand the details of the construction and help to convince oneself that the shape really can be strictly self-assembled and to directly observe how the TAS accomplishes this task.

## 4 Assembly of Finite Shapes

In this section we study the power of nondeterminism in assembling finite shapes. We first show that a finitary analog of Theorem 3.1 holds, by showing that the tile complexity of some shapes can be reduced using nondeterminism. The ideas in this construction will be useful in proving the main theorem of this section, which shows that the minimum tile set problem is $\Sigma_2^\mathsf{P}$-complete.

Recall that all of the TAS's we study are assumed singly-seeded. Let $S \subseteq \mathbb{Z}^2$ be a shape. The *(temperature-2) tile complexity* of $S$ is

$$\mathrm{C}^{\mathrm{tc}}(S) = \min \{\ |T|\ |\ \mathcal{T} = (T, \sigma, 2) \text{ is a TAS and } \mathcal{T} \text{ strictly self-assembles } S\ \},$$

with the convention $\min \varnothing = \infty$. The *(temperature-2) directed tile complexity* of $S$ is

$$\mathrm{C}^{\mathrm{dtc}}(S) = \min \{\ |T|\ |\ \mathcal{T} = (T, \sigma, 2) \text{ is a directed TAS and } \mathcal{T} \text{ strictly self-assembles } S\ \}.$$

We are interested in the problems, given a finite shape, what is its tile complexity, and what is its directed tile complexity? We define two decision problems that are equivalent to these optimization problems. Let $\mathcal{FS} \subset \mathcal{P}(\mathbb{Z}^2)$ denote the set of all finite shapes. The *minimum tile set* problem is

$$\mathrm{MinTileSet} = \{\ \langle S, c \rangle\ \big|\ S \in \mathcal{FS}, c \in \mathbb{Z}^+,\ \text{and } \mathrm{C}^{\mathrm{tc}}(S) \leq c\ \},$$

and the *minimum directed tile set* problem is

$$\mathrm{MinDirectedTileSet} = \left\{\ \langle S, c \rangle\ \bigg|\ S \in \mathcal{FS}, c \in \mathbb{Z}^+,\ \text{and } \mathrm{C}^{\mathrm{dtc}}(S) \leq c\ \right\}.$$

Adleman, Cheng, Goel, Huang, Kempe, Moisset de Espanés, and Rothemund [5] showed that the problem MinDirectedTileSet is NP-complete. In Section 4.2 we show that MinTileSet is $\Sigma_2^\mathsf{P}$-complete, where $\Sigma_2^\mathsf{P} = \mathsf{NP}^{\mathsf{NP}}$. See [7] for a discussion of these complexity classes.

### 4.1 A Finite Shape for which Nondeterminism Reduces Tile Complexity

Although the main result of Section 4, Theorem 4.3, together with the (widely-believed) assumption that $\mathsf{NP} \neq \Sigma_2^\mathsf{P}$ and the fact proven in [5] that MinDirectedTileSet $\in \mathsf{NP}$, implies Theorem 4.2 of this subsection, we prove Theorem 4.2 explicitly in order to illustrate some of the reasoning used in the proof of Theorem 4.3.

Given a shape $S$ (possibly a subshape of a larger shape we wish to self-assemble), we say some tile types *hard-code* $S$ to mean that there are $|S|$ unique tile types, each one specific to a position in $S$, using double glues between tile types of all adjacent positions in $S$.

Given a shape $S$ with a subshape $S' \subset S$, we say $S'$ is an *isolated* subshape of $S$ if there is a point $p \in S'$ such that every path from a point in $S'$ to a point in $S \setminus S'$ includes $p$. In this case, we say $p$ is the *root* of the subshape. If $S'$ is a tree, we say it is an *isolated subtree* of $S$. We say that an isolated subshape $S'$ of $S$ is *singly-connected* if there is precisely one point in $S \setminus S'$ adjacent to the root of $S'$.

**Lemma 4.1.** *Let $S$ be a shape with at least one cycle and $S_T \subseteq S$ be a singly-connected isolated subtree of $S$ with root $r \in S_T$. Any TAS that strictly self-assembles $S$ places at least $\mathrm{C}^{\mathrm{tc}}(S_T)$ unique tile types in $S_T$.*

*Proof.* Let $\mathcal{T}$ be a TAS that strictly self-assembles $S$, and let $\alpha \in \mathcal{A}_\square[\mathcal{T}]$, so that $S_\alpha = S$.

To begin with, we claim that $\alpha(r)$, the tile type on the root $r$, does not appear anywhere else in $S_T$. For the sake of contradiction, suppose there were a position $p \in S_T \setminus \{r\}$ with $\alpha(p) = \alpha(r)$. Since $S_T$ is assumed to be a tree, there is a unique path $\vec{p} = (p_0, \ldots, p_m) \in (\mathbb{Z}^2)^{m+1}$ between $r$ and $p$, such that $p_0 = r$ and $p_m = p$. Let $p'$ be the position in $S \setminus S_T$ that is adjacent to $r$. Depending on whether $r - p' = p - p_{m-1}$ holds or not, we have two cases to be investigated.

The first case is when this equation holds. Note that in this case there must exist a position $p_{m+1} \in S_T$ such that $p_1 - r = p_{m+1} - p_m$. This is because $\alpha(r)$ and $\alpha(p_1)$ are bound via double glue. If the seed of $\mathcal{T}$ is in $S \setminus S_T$, then we can replace the singly-connected subtree of $S_T$ rooted at $p_1$ with the subtree $S_1$ of $S_T$ rooted at $p_{m+1}$. It is impossible that the growth of $S_1$ was blocked in $\alpha$, since $S_1$ is a tree, and this replacement enables it to grow further. Hence, $\mathcal{T}$ could self-assemble a shape strictly smaller than $S$. However, this contradicts that $\mathcal{T}$ strictly self-assembles $S$. This argument works also when $r$ has a third adjacent point, which is in $S_T$, and the seed is in the singly-connected subtree rooted at the point. If the seed is on the path $\vec{p}$ between $r$ and $p$, then $\mathcal{T}$ could repeat this path when the growth reaches $p$, and then continue the self-assembly process after the repetition in the same way as done in the expected assembly at $p$. This growth cannot be blocked by any tile on $S \setminus S_T$ because if it were, then $r$ would be on a cycle in $S$, contradicting the fact that $S_T$ is a subtree of $S$. No tile on $S_T$ can block it either because $S_T$ is a tree. Thus, $\mathcal{T}$ cannot set its seed location on $\vec{p}$. The remaining possibility is when the seed is in the singly-connected subtree rooted at $p$ (let us denote it by $S_2$). Then $\mathcal{T}$ could grow the sub-assembly of the shape $S \setminus S_T$ at $p$ instead of the path reaching to $r$. This alternative assembly process is not blocked by $S_2$. Furthermore, the growth of $S \setminus S_T$ in the expected assembly is not blocked by anything but tiles in $S \setminus S_T$. Thus, the alternative assembly would be strictly smaller than $S$.

Let us consider the other case when the equation $r - p' = p - p_{m-1}$ does not hold. In this case, there must exist points $p'_{m+1}, q \in S_T$ satisfying $r - p' = p - p'_{m+1}$, $r - q = p - p_{m-1}$, and $q \neq p'$. If the seed is in the singly-connected subtree rooted at $p'_{m+1}$, then at $p$, $\mathcal{T}$ could proceed its assembly in a manner expected to occur at $q$ because $\alpha(q)$ can attach to $\alpha(p)$. This results in a terminal assembly strictly smaller than $S$. Otherwise, after reaching $p$, $\mathcal{T}$ could grow the subassembly of shape $S \setminus S_T$ from $p$. This contradicts the fact that $S_T$ is a tree.

This claim has been verified so that $\alpha(r)$ never appears on $S_T \setminus \{r\}$. By replacing the glue of the side that faces $S \setminus S_T$ with a null glue, and furthermore putting the seed on $r$ if the seed of $\mathcal{T}$ is in $S \setminus S_T$, then we can construct a new TAS $\mathcal{T}'$ that strictly self-assembles $S_T$ without changing any tiles on $S_T \setminus \{r\}$. Thus any TAS which strictly self-assembles $S$ needs at least $\mathrm{C}^{\mathrm{tc}}(S_T)$ tile types to assemble $S_T$. □

If $S$ is a tree, the analogous result of this lemma does not hold. Let us consider a tree $U = \{(0,2), (0,1), (0,0), (1,0), (2,0), (2,1), (2,2)\}$ and let $J = U \setminus \{(0,2)\}$. Easily we can see that $\mathrm{C}^{\mathrm{tc}}(J) = 6$, while $\mathrm{C}^{\mathrm{tc}}(U) = 5$.

**Theorem 4.2.** *There is a finite shape $S \subset \mathbb{Z}^2$ such that $\mathrm{C}^{\mathrm{tc}}(S) < \mathrm{C}^{\mathrm{dtc}}(S)$.*

*Proof.* The shape $S$ is shown in Figure 2. In the following, the loop $L$ means the shape which consists of $L_0, L_1, \ldots, L_h, L'_0, L'_1, \ldots, L'_h$, the tile between $L_h$ and $L'_h$, and the tile between $L_0$ and $L'_0$.

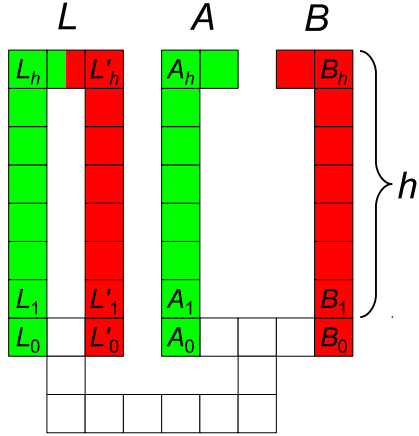

Figure 2: A finite shape $S$ for which $\mathrm{C}^{\mathrm{tc}}(S) < \mathrm{C}^{\mathrm{dtc}}(S)$. Nondeterminism is forced to occur at the two-color top-middle position of the loop $L$, since any minimal tile set must reuse the tile types from subtrees $A$ and $B$ to create $L$.

The height $h$ is left as a variable parameter; increasing $h$ increases the gap between $\mathrm{C}^{\mathrm{tc}}(S)$ and $\mathrm{C}^{\mathrm{dtc}}(S)$. Let us index the tile positions on the pillar $A$ from its bottom as $A_1, A_2, \ldots, A_h$ and do the same for $B$ as $B_1, B_2, \ldots, B_h$. In a similar manner, the left and right pillars of the loop $L$ are indexed as $L_1, L_2, \ldots, L_h$ and $L'_1, L'_2, \ldots, L'_h$, respectively.

First we establish that $\mathrm{C}^{\mathrm{tc}}(S) \leq 2h + 16$ (actually with equality, but we only require and only prove an upper bound). If the seed is placed in the bottom row, 12 tile types (including the seed) hard-code the gray positions, $h + 2$ tile types hard-code the subtree $A$, and $h + 2$ tile types hard-code the subtree $B$. The tile types at $A_0, \ldots, A_h$ can be reused at $L_0, \ldots, L_h$, and the tile types at $B_0, \ldots, B_h$ can be reused at $L'_0, \ldots, L'_h$. Note that this TAS is not directed because the top-middle position of the loop could receive either a tile from $A$ or a tile from $B$. Furthermore, these must be different tile types, because the top-right tile type in $A$ must have a double glue on its west but cannot have a double glue on any other side, whereas the top-left tile type of $B$ must have a double glue on its east but not on any other side.

We now show that this nondeterminism is necessary to achieve minimum tile complexity. In particular, we will show that $\mathrm{C}^{\mathrm{dtc}}(S) \geq 3h$. Let $S_T$ be the tree which consists of the pillars $A, B$ and the three tiles connecting them. Let $\mathcal{T}$ be a directed TAS that strictly self-assembles $S$. Lemma 4.1 implies that any TAS that strictly self-assembles $S$ needs $\mathrm{C}^{\mathrm{tc}}(S_T)$ tile types to assemble $S_T$, and due to Theorem 4.3 in [5], $\mathrm{C}^{\mathrm{tc}}(S_T) = 2h + 5$.

First we consider the case when $\mathcal{T}$ places its tiles such that every pair of adjacent tiles on the loop $L$ is bound via double glue. Being singly-seeded, either the left or the right pillar of the loop $L$ does not contain the seed; assume without loss of generality that the left pillar does not. Since all tiles on the loop are double-bonded, the left pillar can grow upward as $L_1 \to L_2 \to \cdots \to L_h$. Note that the tile on the top-middle position of the loop $L$ should be different from the one on the end of the pillar $A$; otherwise the gap between the ends of $A$ and $B$ would be filled with the tile on $L'_h$. This means that the left pillar cannot reuse the tiles on $A$, and trivially it cannot reuse the ones on $B$. Thus, in this case, $\mathcal{T}$ contains at least $3h$ tile types.

Next we consider the case when some of adjacent tiles on the loop $L$ are not bound via double glue. Note that at most 2 such weak bonds can appear on the loop, and furthermore, they must be incident on a single tile. Thus, we assume without loss of generality that such a weak bond does not occur on the left pillar. Depending on whether the bond between $L'_h$ and the tile on the top-middle of $L$ is weak or not, there are two subcases to be investigated. If it is weak and the seed is not on the right pillar, a similar argument as above enables us to see that any tiles on $A$ or $B$ cannot be reused for the right pillar (the right pillar must grow upward because of the weak bond). If the bond is weak and the seed is on $L'_i$, then on $L'_{i+1}, \ldots, L'_h$, $\mathcal{T}$ cannot put any tile placed on $A$ or $B$. Furthermore, if $\mathcal{T}$ reuses some tile from $A$ or $B$ and places it on some of $L'_1, \ldots, L'_i$, then the bottom row could extend to the left of the loop into empty space. Finally, we consider the second subcase when $L'_h$ binds to the tile on the top-middle position via double glue, but $L'_j$ does not bind to $L'_{j+1}$ via double glue for some $1 \leq j < h$. This establishes that the left pillar must be hardcoded by $h$ new tile types, whence $\mathcal{T}$ contains at least $3h$ tile types. □

### 4.2 The Minimum Tile Set Problem is $\Sigma_2^P$-complete

The following is the main theorem of Section 4.

**Theorem 4.3.** MINTILESET *is $\Sigma_2^P$-complete.*

*Proof.* To show that MINTILESET $\in \Sigma_2^P$, define the verification language

$$\text{MINTILESET}_V = \left\{ \langle S, c, \mathcal{T}, \vec{\alpha} \rangle \;\middle|\; \begin{array}{l} S \in \mathcal{FS}, c \in \mathbb{Z}^+, \mathcal{T} = (T, \sigma, 2) \text{ is a TAS with} \\ |T| \leq c, \vec{\alpha} = (\sigma, \alpha_2, \alpha_3, \ldots, \alpha_k) \text{ is a } \mathcal{T}\text{-assembly} \\ \text{sequence with } S_{\alpha_k} = S, \text{ and } \alpha_k \text{ is } \mathcal{T}\text{-terminal} \end{array} \right\}.$$

Clearly MINTILESET$_V \in$ P. MINTILESET $\in \Sigma_2^P$ because $\langle S, c \rangle \in$ MINTILESET if and only if there exists $\mathcal{T} = (T, \sigma, 2)$ with $|T| \leq c$ such that for all $\mathcal{T}$-assembly sequences $\vec{\alpha} = (\sigma, \alpha_2, \ldots, \alpha_k)$ of length $k = |S|$, $\langle S, c, \mathcal{T}, \vec{\alpha} \rangle \in$ MINTILESET$_V$, with $|\langle \mathcal{T} \rangle|$ and $|\langle \vec{\alpha} \rangle|$ bounded by $O(|\langle S, c \rangle|^2)$.

To show that MINTILESET is $\Sigma_2^P$-hard, we show that $\exists\forall$CNF-UNSAT $\leq_m^P$ MINTILESET, where $\exists\forall$CNF-UNSAT is the $\Sigma_2^P$-complete language [37, 40, 45]

$$\exists\forall\text{CNF-UNSAT} = \left\{ \langle \varphi \rangle \;\middle|\; \begin{array}{l} \varphi \text{ is a true quantified Boolean formula } \varphi = \exists x \forall y \neg \phi(x, y), \\ \text{where } \phi \text{ is an unquantified CNF formula with } n + m \\ \text{input bits } x = x_1, \ldots, x_n \text{ and } y = y_1, \ldots, y_m \end{array} \right\}.$$

We follow a similar strategy to the reduction of 3SAT to MINDIRECTEDTILESET shown in [5]. The $\leq_m^P$-reduction $\langle \varphi \rangle \mapsto \langle S, c \rangle$ works as follows. First, we compute a tree $\Upsilon \in \mathcal{FS}$ that "represents" $\varphi$ with subtree gadgets that encode possible variable assignments and their effect on clauses. We then process $\Upsilon$ with the polynomial-time algorithm described in [5] that computes the minimum number of tile types needed to strictly self-assemble a tree. Let $\mathcal{T} = (T, \sigma, 2)$ be this minimal TAS that strictly self-assembles $\Upsilon$, and let $c = |T|$. We then compute a shape $S \in \mathcal{FS}$ such that $\Upsilon \subset S$ with the property that, if $\varphi$ is true, then the tile types in $T$ can be modified, solely through changing some null glues to be single or double glues, producing a TAS $\mathcal{T}' = (T', \sigma, 2)$ with $|T'| = |T| = c$ such that $\mathcal{T}'$ strictly self-assembles $S$, and if $\varphi$ is false, then no TAS with at most $c$ tile types can strictly self-assemble $S$. The shape $S$ is shown in Figure 3. In Figure 3, the height of pillars is set to a number bigger than $20\ell$, where $\ell$ is the number of variables in $\varphi$.[9]

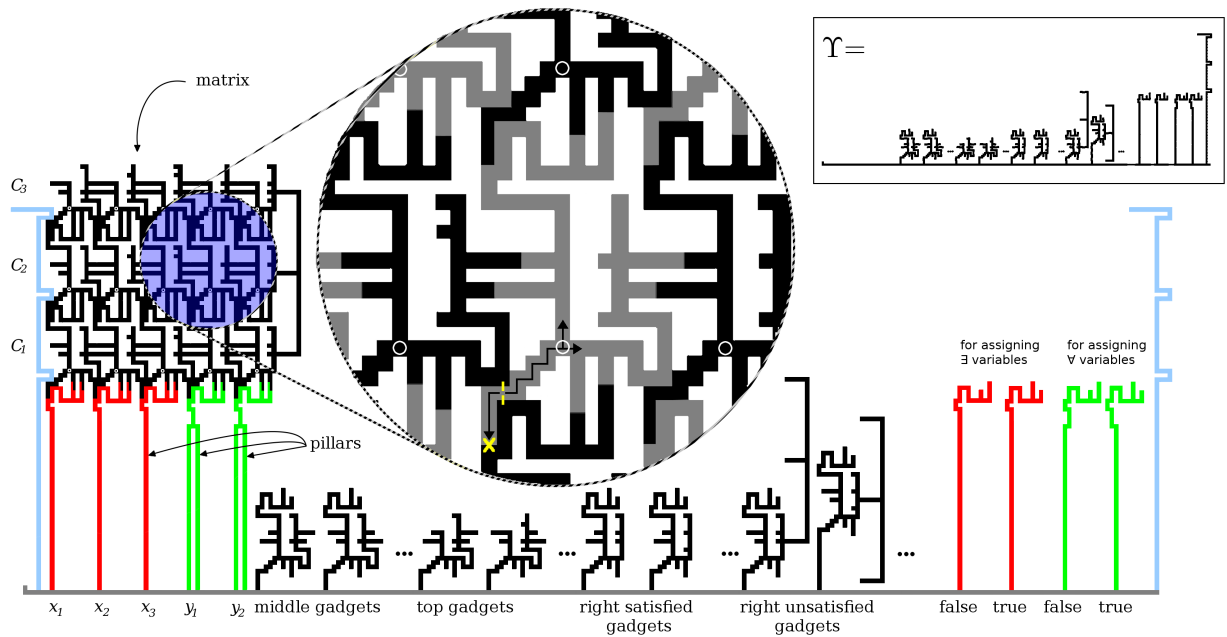

Figure 3: The shape $S$ of the reduction $\langle \varphi \rangle \mapsto \langle S, c \rangle$ showing $\exists\forall$CNF-Unsat $\leq_m^{\mathsf{P}}$ MinTileSet. In this example, the quantified negated CNF formula $\varphi = \exists x \forall y \neg \phi(x, y)$ has clauses $C_1, C_2$ and $C_3$, $\exists$-variables $x_1, x_2$, and $x_3$, and $\forall$-variables $y_1$ and $y_2$. The "matrix" of gadgets at the top left has a row of gadgets for each clause and a column of gadgets for each variable. The matrix sits atop a group of "pillars" that, when tiled by actual tiles, will represent a variable assignment to $\phi$ (along with one taller left-boundary pillar to help initiate cooperative binding of gadgets to assemble the matrix). The tree $\Upsilon$ is $S$ without the matrix and pillars beneath it. In the zoom-in, the two yellow lines above the yellow X represent strength-1 glues that cooperate to place the gray gadget once (enough of) the black gadgets to its west and south are in place. The yellow X shows "backward growth" of the gray gadget that is blocked before it can grow down far enough to form a new copy of the bottom row of $S$.

Suppose that $\phi$ has $k$ clauses $C_1, \ldots, C_k$ and $\ell = n + m$ input variables $v_1, \ldots, v_\ell$, where $v_1, \ldots, v_n = x_1, \ldots, x_n$ are the $\exists$-variables of $\varphi$ and $v_{n+1} \ldots, v_\ell = y_1, \ldots, y_m$ are the $\forall$-variables of $\varphi$. A clause $C$ is *satisfied* by variable $v$ if $C$ contains literal $v$ and $v$ is true, or if $C$ contains literal $\neg v$ and $v$ is false. For each $1 \leq i \leq k$ and $1 \leq j \leq \ell$, define the following six gadgets:

1. SST$_{ij}$: $C_i$ is satisfied by $v_p$ for some $1 \leq p < j$, and $v_j$ is true.

---

[9]Actually, it is enough to set the height of pillars to any number bigger than the width of the clause-variable matrix.

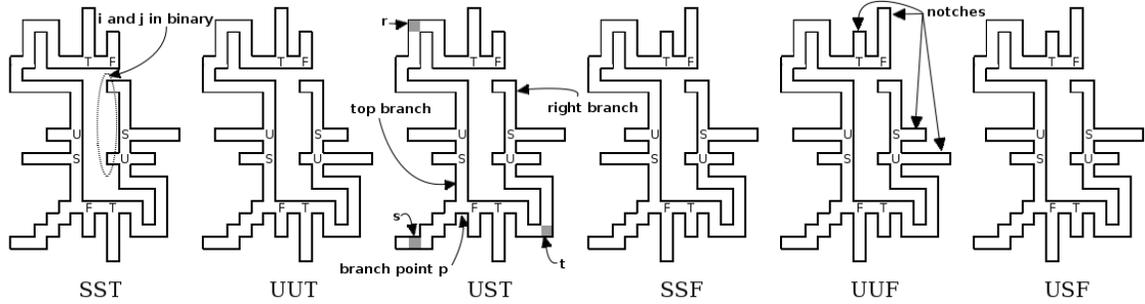

Figure 4: Six main varieties of "information-bearing" tree gadgets used in the reduction. The position $(i,j)$ where the gadget is intended to go in the matrix is encoded in binary. Gadgets intended for the top row are missing the top "T/F" bumps, and gadgets intended for the right column are different on the right depending on whether the clause is satisfied or not, as shown in Figure 3.

2. $\text{SSF}_{ij}$: $C_i$ is satisfied by $v_p$ for some $1 \leq p < j$, and $v_j$ is false.

3. $\text{UUT}_{ij}$: $C_i$ is unsatisfied by $v_p$ for every $1 \leq p \leq j$, and $v_j$ is true.

4. $\text{UUF}_{ij}$: $C_i$ is unsatisfied by $v_p$ for every $1 \leq p \leq j$, and $v_j$ is false.

5. $\text{UST}_{ij}$: $C_i$ is unsatisfied by $v_p$ for every $1 \leq p < j$, $C_i$ is satisfied by $v_j$, and $v_j$ is true.

6. $\text{USF}_{ij}$: $C_i$ is unsatisfied by $v_p$ for every $1 \leq p < j$, $C_i$ is satisfied by $v_j$, and $v_j$ is false.

Each of these six main varieties of "information-bearing" gadgets is shown in Figure 4. Each gadget is designed to minimize the amount of "potential unwanted cooperative strength-1 binding" when they are placed next to each other in the "matrix" of gadgets in the upper left of Figure 3.[10] Each gadget encodes the integers $i$ and $j$, as well as encoding the information about the clause $C_i$ and variable $v_j$ as described above. Some of the "boundary case" gadgets are shaped slightly differently than those in Figure 4. If $i = k$ (a "top gadget"), the top of the gadget will not encode information about the truth value of the variable $v_i$. If $j = \ell$ (a "right gadget"), the right side of the gadget will still encode whether the clause is satisfied, but the gadget will have a different shape than for $1 \leq j < \ell$. These special boundary shapes are shown in Figure 3.

Not all six varieties of gadgets are created for each $(i,j)$; the only gadgets created are those that are logically consistent with some variable assignment to $\phi$. The matrix and pillars portion of $S$ (i.e., $S \setminus \Upsilon$) depends only on the number of $\exists$-variables, the number of $\forall$-variables, and the number of clauses. The remainder of the information about $\varphi$ is encoded in the following choices about which gadgets to create in $\Upsilon$. In the case of $j = 1$, the gadgets $\text{SST}_{i1}$ and $\text{SSF}_{i1}$ are not created. For any clause $C_i$ in which the literal $v_j$ (resp. $\neg v_j$) does not appear, the gadget $\text{UST}_{ij}$ (resp. $\text{USF}_{ij}$) is not created. Similarly, for any clause $C_i$ in which *no* literal $v_p$ (resp. $\neg v_p$) appears for any $1 \leq p < j$, the gadget $\text{SST}_{ij}$ (resp. $\text{SSF}_{ij}$) is not created. Finally, for any clause $C_i$ in which the literal $v_j$ (resp. $\neg v_j$) *does* appear, the gadget $\text{UUT}_{ij}$ (resp. $\text{UUF}_{ij}$) is not created.

---

[10]Strength-1 glues can only have an effect on growth of gadgets in the matrix when they are on tiles on the gray positions $r$, $s$, and $t$ in Figure 4, if the tile types used to assemble those gadgets in the matrix are the same as those used to assemble the gadgets in $\Upsilon$. This is useful in proving the converse direction of the reduction by showing that if a tile assembly system with $\leq c$ tile types strictly self-assembles $S$, then $\varphi$ must be true.

The tree $\Upsilon$ is $S$ without the "matrix" on the top left and the "pillars" beneath it that connect it to the bottom row. Let $c = \text{C}^{\text{tc}}(\Upsilon)$. We assume that the seed is placed on the rightmost position of the bottom row, for both the shapes $\Upsilon$ and $S$. At the end of the proof we show how to modify the shapes to enforce this restriction. The steps needed to complete the proof are divided into several lemmas. These lemmas are proven after the current proof. Lemmas 4.4 and 4.7 establish each direction of the claim that $\varphi$ is true $\iff \text{C}^{\text{tc}}(S) \leq c$. Intuitively, since $\Upsilon$ is a "tree-like" subshape of $S$ (despite the leftmost tiles intersecting cycles in $S$), any tile system that strictly self-assembles $S$ must place tiles in the bottom row that do not appear anywhere else in $\Upsilon$. $\varphi$ is true $\implies \text{C}^{\text{tc}}(S) \leq c$ because we can modify the null glues of tiles in the left half of the bottom row of $\Upsilon$ to be double glues matching those tile types from the pillars on the right to grow the pillars on the left. In the case of the $\exists$-variables $x$ we choose an assignment by our choice of double glues. In the case of $\forall$-variables $y$ we have no choice; we must allow both the "false" and "true" pillars to grow and nondeterministically compete to assign a bit to each $y_i$. We can then modify null glues in the gadgets and pillars to be single glues that propagate information about the neighbors of a gadget to allow a new gadget encoding the proper information to be placed in the matrix. Therefore the assembly of the matrix "evaluates $\phi(x, y)$" and if it is false, strictly self-assembles $S$. The reverse direction is more tedious to establish. Again, since $\Upsilon$ is a "tree-like" subshape of $S$, any TAS strictly self-assembling $S$ already uses $c$ tile types just to assemble the $\Upsilon$ portion of $S$ (derived from Lemma 4.5). Therefore to assemble all of $S$ using $c$ tile types requires reusing these same tile types. Our gadget design, together with the properties of minimal tile sets for trees, allow us to conclude that the only way to tile the matrix is "using the gadgets in the way they were intended", which means the rightmost vertical bar of the matrix cannot form unless at least one clause is not satisfied; i.e., $\varphi$ is true.

To handle the placement of the seed, define $a \in S$ to be the rightmost point on the bottom row of $S$. Make two copies of $S$, place one directly above the other but without touching, and connect the copies by a width-1 "bridge" of length $h$ that connects to each copy of $a$ on $a$'s right side. Denote this new shape by $S'$. It is routine to show using techniques similar to those in the proof of Lemma 4.1 that any minimal TAS for $S'$ uses $\text{C}^{\text{tc}}(S) + h$ tile types, places the seed in the bridge, uses $h$ tile types to grow the bridge and uses the tile types of a TAS $\mathcal{T}$ that is minimal (subject to the restriction that $\mathcal{T}$ places the seed at $a$) for $S$, to assemble each copy of $S$. Let $c' = c + h$. Our reduction outputs $\langle S', c' \rangle$, rather than $\langle S, c \rangle$. By the arguments above concerning $S$, we have that $\varphi$ is true $\iff \text{C}^{\text{tc}}(S') \leq c'$, whence $\exists\forall\text{CNF-UNSAT} \leq_{\text{m}}^{\text{P}} \text{MINTILESET}$. □

In the following lemmas, $\varphi$ denotes an arbitrary (true or false) quantified Boolean formula of the form $\varphi = \exists x \forall y \neg \phi(x, y)$, where $\phi$ is an unquantified CNF formula with $n + m$ input bits $x = x_1, \ldots, x_n$ and $y = y_1, \ldots, y_m$. $\Upsilon, S \in \mathcal{FS}(\mathbb{Z}^2)$ refer to the tree and shape constructed from $\varphi$ as in the proof of Theorem 4.3, and $c = \text{C}^{\text{tc}}(\Upsilon)$.

**Lemma 4.4.** *If $\varphi$ is true, then $\text{C}^{\text{tc}}(S) \leq c$.*

*Proof.* Let $\mathcal{T}_\Upsilon = (T, \sigma, 2)$ be a minimal TAS that strictly self-assembles $\Upsilon$ with seed placed at position $a$, the rightmost point of the bottom row of $\Upsilon$, and let $\alpha \in \mathcal{A}_\square[\mathcal{T}_\Upsilon]$ be the unique terminal producible assembly of $\mathcal{T}_\Upsilon$, such that $S_\alpha = \Upsilon$. Theorem 4.3 of [5] shows that if $\mathcal{T}_\Upsilon$ is a minimal TAS for $\Upsilon$, then $\mathcal{T}_\Upsilon$ puts the same tile type in two positions $p_1, p_2 \in \Upsilon$ if and only if the subtrees of $\Upsilon$ rooted at $p_1$ and $p_2$ (with the seed location considered the root of $\Upsilon$) are isomorphic and "identically entered" (meaning both of them have their parent in the same direction). This theorem is stated for directed TAS's but it is easy to show that any minimal TAS for a tree must be directed. Given $\alpha$ and $t \in T$, define $t$ to be *singular in $\alpha$* if it appears exactly once in $\alpha$. Thus, all the positions at the bottom row of $\Upsilon$ will receive tile types that are singular in $\alpha$.

Since $\varphi$ is true, there is an assignment $f$ to variables $x_1, x_2, \ldots, x_n$ such that any assignment to $y_1, y_2, \ldots, y_m$ makes $\phi(x, y)$ false. Since all the tile types in the bottom row are unique, we can change the north glues of the tiles at the base of the clause-variable matrix, without ruining the rest of the shape. We change the north glues so that the blue pillar grows as the leftmost pillar and for each variable $x_i$ we grow the red true/false pillar depending on $f(x_i)$ being true or false. For the positions corresponding to $y$ variables, we change the north glues so that both true and false green pillars can grow.

We will also change a number of null glues into single glues in the following way: the set of labels with strength one will be T, F, $S_{i,j}$, and $U_{i,j}$, where $1 \leq i \leq k$ and $1 \leq j \leq \ell$. For each gadget $G$, let $p$ be the position of its branch point in $\Upsilon$, as shown in Figure 4. Then, we change the null glue of the south side of $\alpha(s)$ to a single glue with label T if $G$ is of type SST, UUT, or UST; otherwise, we change the south glue to a single glue with label F. Also, we change the north glue of $\alpha(s)$ to a single glue with label $S_{i,j}$ if $G$ is a gadget for the $i$th clause and $j$th variable and is of type SST or SSF; otherwise, if $G$ is of type UUT, UUF, UST, or USF, the north glue of $\alpha(s)$ will be a single glue with label $U_{i,j}$. Note that, the tile type $\alpha(s)$ is singular in $\alpha$, due to the fact that its rooted subtree contains the encoding about the gadget type, clause number, and variable number, and hence, is not isomorphic to any other subtree.

The north glue of the tile type $\alpha(r)$ will be changed to a strength one glue with label T if $G$ is of type SST, UUT, or UST; otherwise, its label will be F. The south glue of the tile type $\alpha(t)$ will be changed to a strength one glue with label $S_{i,j+1}$ if $G$ is for the $i$th clause and $j$th variable and is of type SST, SSF, UST, or USF; otherwise, its label will be $U_{i,j+1}$.

Applying the above-mentioned changes will give us a TAS $\mathcal{T}_S$ with the same tile complexity as $\mathcal{T}_\Upsilon$. In $\mathcal{T}_S$, a gadget can grow at the cell of the matrix corresponding to clause $i$ and variable $j$ using cooperation of single glues of the bottom and left gadget if and only if its notches match the notches of the bottom and left gadget. And, the notches of gadgets are designed in a way that they can be put together to assemble $S$ if and only if the truth assignment to $x$ and $y$ variables (presented as pillar notches at the first row of the matrix) make $\phi(x, y)$ false. Intuitively, the gadgets grow in the matrix so as to "evaluate $\phi(x, y)$" on inputs $x$ and $y$ encoded in the pillars, with the pillars encoding $y$ nondeterministically choosing values for each of the $y_i$. If we choose the proper assignment $f$ to $x$, such that, for all assignments to $y$ (corresponding to different terminal assemblies), $\phi(x, y)$ is false (such an assignment $f$ exists since $\varphi = \exists x \forall y \neg \phi(x, y)$ is true), then all of these assemblies will have at least one unsatisfied right gadget in the rightmost column of the matrix, and the assembly will have shape $S$. Therefore $\mathcal{T}_S$ strictly self-assembles $S$. □

From here until the end of the section, assume that $\mathcal{T}_S = (T_S, \sigma, 2)$ is a TAS that strictly self-assembles $S$ with the seed placed at the rightmost position on the bottom row. Also, $B$ represents the subshape of $S$ that does not have the black matrix on the left, but has the pillars beneath it, and $I \subset \mathbb{Z}^2$ denotes the set of $k \times \ell$ positions (where $k$ is the number of clauses in $\phi$ and $\ell$ is the number of variables) marked by small circles in Figure 3. A set $I' \subseteq I$ is called a *staircase* if the following implication holds:

$$\big[(x_1, y_1) \in I', (x_2, y_2) \in I, x_2 \leq x_1, \text{ and } y_2 \leq y_1\big] \implies (x_2, y_2) \in I'.$$

Let $G^* = \bigcap_{G \in \mathcal{G}} G$, where each element of $\mathcal{G}$ is a set of tile positions of a gadget created in the proof of Theorem 4.3 (Figure 4) translated so that the branch tile is at the origin (so, $\mathcal{G}$ has at most $k \times \ell \times 6$ elements).

**Lemma 4.5.** *If $|T_S| \leq c = \mathrm{C}^{\mathrm{tc}}(\Upsilon)$, then there is a TAS $\mathcal{T}_\Upsilon = (T_\Upsilon, \sigma, 2)$ that can be obtained from $\mathcal{T}_S = (T_S, \sigma, 2)$ by only changing a number of double glues to null glues (in particular implying that $|T_\Upsilon| = |T_S|$), such that $\mathcal{T}_\Upsilon$ strictly self-assembles $\Upsilon$.*

*Proof.* It suffices to show that if $\alpha \in \mathcal{A}_\square[\mathcal{T}_S]$, then for any two adjacent positions $p_1, p_2 \in \Upsilon$, $\alpha(p_1)$ and $\alpha(p_2)$ are bound together by a double glue. This will establish that, by adjusting double glues on the north of tiles beneath the pillars to be null glues, all of $\Upsilon$ can grow from the tiles without growing any of the pillars or matrix of $S - \Upsilon$. Because each adjacent tile in the row under the matrix double bonded, each of these tile types must be singular in $\alpha \upharpoonright \Upsilon$ ($\alpha$ restricted to $\Upsilon$); otherwise, their appearance elsewhere in $\alpha$ would lead to copies of the pillars and matrix growing in a second location.

For the sake of contradiction, let $p_1 \in \Upsilon$ and $p_2 \in \Upsilon$ be the closest positions to the seed that are adjacent to each other but $\alpha(p_1)$ and $\alpha(p_2)$ do not have double glue between them. Then, they are on the bottom row below the clause-variable matrix in $S$. Thus, there must be a pillar growing down from the clause-variable matrix; consider the pillar that grows down in $\alpha$ the earlier than the others. This pillar cannot reuse any tile type that is used in positions to the right of $p_1$ in $\Upsilon$; otherwise, an undesirable part of $\Upsilon$ can grow to the left of the downward pillar. Therefore, the number of tile types in $\mathcal{T}_S$ is at least $\mathrm{C}^{\mathrm{tc}}(\Upsilon) - x + y$, where $x$ is the horizontal width of the clause-variable matrix and $y$ is the height of the pillars. This is a contradiction to the assumption that $\mathcal{T}_S$ uses at most $\mathrm{C}^{\mathrm{tc}}(\Upsilon)$ tile types, since we set $y > x$ in our construction. □

The following lemma states the "inductive step" of the proof of Lemma 4.7. Namely, if gadgets of a minimal tile system for $S$ grow to fill in part of the matrix "in the way we intend", then the only way to extend this growth to fill in an additional gadget is also "in the way we intend."

In the following lemma, "right branch" and "top branch" refer to the two subtrees of a gadget rooted at the branch as shown in Figure 4.

**Lemma 4.6.** *Suppose $\mathcal{T}_S$ has at most $c = \mathrm{C}^{\mathrm{tc}}(\Upsilon)$ tile types. Let $\alpha \in \mathcal{A}[\mathcal{T}_S]$ be a producible assembly such that*

1. *$B \subseteq S_\alpha \subseteq S$. (where $B$ is the subshape of $S$ that does not have the black matrix on the left, but has the pillars beneath it)*

2. *$S_\alpha \cap I$ is a staircase of cardinality $m < |I|$.*

3. *All tile types in $\alpha(S_\alpha \cap I)$ are branch tiles of gadgets.*

4. *The right and top branches of tiles in $S_\alpha \cap I$ are present in $\alpha$.*

5. *If there exists $p \in I$ such that $p \notin S_\alpha$, then $S_\alpha \cap (p + G^*) = \varnothing$.*

*Then there exists an assembly $\beta \in \mathcal{A}[\mathcal{T}_S]$ such that $\alpha \to \beta$ and requirements (1)-(5) are satisfied with "$\alpha$" replaced by "$\beta$" and "$m$" replaced by "$m+1$".*

*Proof.* $\alpha$ cannot be a terminal assembly, since $\mathcal{T}_S$ strictly self-assembles $S \neq S_\alpha$.

Since double glues cannot be added to gadget tile types without ruining the tree shape portion of $S$, $\alpha$ must grow by cooperation of single glues. This cooperation can happen only at $F - (4,3)$, where
$$F = \{p \in I \setminus S_\alpha : (S_\alpha \cap I) \cup \{p\} \text{ is a staircase}\}.$$

In other words, $F - (4,3)$ is the set of points marked $s$ in Figure 4, which are adjacent to points marked $r$ and $t$ in neighboring gadgets. Let $\beta'$ be a minimal assembly producible from $\alpha$ such that $S_{\beta'} \cap F$ is not empty. Let $\{p\} = S_{\beta'} \cap F$. All paths from $S_\alpha$ to $p$ in $\beta'$ must pass through $s = p - (4,3)$, due to the minimality of $\beta'$. Even if the tile type that goes in $s$ is not taken from any gadgets, the tile type that eventually goes to $s + (1,1)$ must be chosen from a gadget, since it should be able to grown the zig-zag shape only by double glues, and the zig-zag shape is used only in one of the gadgets. Thus, $\beta'(p)$ is a branch tile type.

Since $p \in S_{\beta'} \cap I$ and $p \notin S_\alpha \cap I$, $S_{\beta'} \cap I$ has cardinality at least $m + 1$. Let $\beta$ be the minimal assembly producible from $\beta'$ in which the right and top branches of $p$ are tiled. $\beta(p)$ and $\beta(q)$ for all $q$ in the branches of $p$ must be tile types from the correct gadget to ensure consistency with the notches of neighboring gadgets and consistency with the (row,column) identifier notches at the top of the right branch.

Due to the minimality of $\beta$, it satisfies condition 5. □

**Lemma 4.7.** *If $\mathcal{T}_S$ has at most $c$ tile types, then $\varphi$ is true.*

*Proof.* First we show that $B$ is $\mathcal{T}_S$-producible. According to Lemma 4.5, by changing a number of double glues in $\mathcal{T}_S$ to null glues, we can obtain a TAS $\mathcal{T}_\Upsilon$ that strictly self-assembles $\Upsilon$. So, $\Upsilon$ is $\mathcal{T}_S$-producible. Moreover, as can be checked in the proof of Lemma 4.5, the null glues in $\mathcal{T}_\Upsilon$ that are double glues in $\mathcal{T}_S$ are the north glues of the tile types that appear at the base of pillars beneath the gadget matrix. Also, all the pillars must grow from the bottom row to the matrix, and not downward, because growing a pillar downward requires adding a double glue to a tile type in the matrix area, which will also ruin $\Upsilon$. Thus, $B$, which is $\Upsilon$ together with the pillars beneath the matrix, is $\mathcal{T}_S$-producible. This establishes the base case.

Let $f(x_i)$ be true if the red true pillar is used to grow the pillar corresponding to $x_i$ and be false if the red false pillar is used. Using Lemma 4.6 for the inductive case, we conclude that there is an assembly $\alpha$ such that $B \cup I \subseteq S_\alpha$ and valid gadgets are/can be used to fill the matrix part of $\alpha$. By our construction of gadgets, this implies that the truth assignment $f$ to $x$ makes $\phi(x,y)$ false for every value of $y$. Thus $\varphi$ is true. □

## 5 Conclusion

We have investigated the power of nondeterminism for the strict self-assembly of shapes in the abstract Tile Assembly Model. We showed that for both the infinite and finite cases, even when the shape is required to be strictly self-assembled, nondeterminism can help to assemble the shape, by making strict self-assembly possible in the infinite case, and reducing tile complexity in the finite case. Furthermore, the problem of finding the minimum tile set to strictly self-assemble a shape is strictly harder (in the sense of nondeterministic time complexity) than that of finding the minimum directed tile set that does so, unless $\mathsf{NP} = \Sigma_2^\mathsf{P}$.

There are some interesting questions that remain open:

1. What is the fastest growing function $f : \mathbb{N} \to \mathbb{N}$ for which one could prove a statement of the form "*For infinitely many $n \in \mathbb{N}$, there is a finite shape $S \subset \mathbb{Z}^2$ such that $\mathrm{C}^{\mathrm{tc}}(S) \leq n$ and $\mathrm{C}^{\mathrm{dtc}}(S) \geq f(n)$*"? The proof of Theorem 4.2 of the present paper establishes this statement for $f(n) = 1.4999n$. Can $f(n)$ be made, for example, $n^2$ or $2^n$? What is an upper bound for $f$ above which such a statement is false? Note that Theorem 3.1 establishes such a statement for *all* functions $f : \mathbb{N} \to \mathbb{N}$ if the shape is allowed to be infinite. However, when designing complex tile systems, a common challenge is to direct a group of tiles to stop growing,[11] so it would be interesting to identify a family of *finite* shapes with a fast-growing gap between the two tile complexity measures. This would imply that sometimes it *really* helps to employ nondeterminism.

2. We have showed that the optimization problem of finding precisely the smallest number of tile types to strictly self-assemble a shape is $\Sigma_2^{\mathsf{P}}$-hard. Can it be shown that for some $\alpha > 1$, the solution is $\Sigma_2^{\mathsf{P}}$-hard to approximate within multiplicative factor $\alpha$? This may be related to Question 1.

3. Is there an $\alpha > 1$ such that it is NP-hard to find an $\alpha$-approximate solution to the minimum directed tile set problem?

4. Shape-building is one common goal of self-assembly; pattern-painting is another. In particular, it is possible to assemble some patterns, such as disconnected sets, if we change the definition of what is interpreted as the assembled object. We say that a TAS $\mathcal{T} = (T, \sigma, 2)$ *weakly self-assembles* a set $S \subseteq \mathbb{Z}^2$ if there is a subset $B \subseteq T$ (the tile types that are "painted black") such that, for all $\alpha \in \mathcal{A}_\square[\mathcal{T}]$, $\alpha^{-1}(B) = S$. In other words, the set of positions with a black tile is guaranteed to be $S$. In the case $B = T$, this definition is equivalent to strict self-assembly, but for $B \subsetneq T$ the shape is allowed to grow outside the desired pattern using tile types from $T \setminus B$ to allow "extra computation room" for painting the pattern using tile types from $B$. Such a definition is appropriate for modeling practical goals such as self-assembled circuit layouts [21, 25, 28, 31, 32, 46] or placement of guides for walking molecular robots [26]; see [22, 23] for more discussion of the theoretical issues of weak self-assembly.[12] It remains open to prove or disprove analogs of Theorems 3.1 and 4.2, with "weakly" substituted for "strictly". In other words, is it possible to uniquely paint an infinite (resp. finite) pattern with a tile system, but every tile system that does so (resp. that does so with no extra tile types) is not directed?

5. It remains open to prove or disprove analogs of Theorems 3.1 and 4.2, with "weakly" substituted for "strictly" *and* with "strict" substituted for "directed". In other words, is it possible to uniquely paint an infinite (resp. finite) pattern with a tile system, but every tile system that does so (resp. that does so with no extra tile types) must self-assemble more than one shape on which this pattern is painted?

---

[11]For example, $\mathrm{C}^{\mathrm{tc}}(S) \approx \mathrm{C}^{\mathrm{dtc}}(S) = O(\log n / \log \log n)$ for $S$ an $n \times k$ rectangle with $n \geq k \geq \log n / \log \log n$, but $\mathrm{C}^{\mathrm{tc}}(S)$ and $\mathrm{C}^{\mathrm{dtc}}(S)$ increase steadily towards $n$ as $k$ decreases from $\log n / \log \log n$ to 1; "counting" to the length of the rectangle and then stopping becomes more difficult as the rectangle's width decreases.

[12]In contrast to the case for strict self-assembly, it can be shown that it is uncomputable to determine the minimum size tile assembly system that weakly self-assembles a given finite shape. This follows from a "Berry's paradox" argument, similar to the one used to show that Kolmogorov complexity is uncomputable, together with the fact that arbitrary algorithms may be simulated in a tile assembly system. Briefly, assuming this quantity is computable, define a Turing machine $M$ that on input $c \in \mathbb{Z}^+$ enumerates finite sets of points lying entirely on the positive $x$-axis until a set $S(c)$ is found whose "weak self-assembly tile complexity" exceeds $c$. Then for each $c \in \mathbb{Z}^+$ define a tile system $\mathcal{T}$ that simulates $M(c)$ in the second quadrant, using its output to place black tiles precisely on points in $S(c)$. Since $\mathcal{T}$ requires only $\log c + O(1)$ tile types, for sufficiently large $c$ this contradicts the tile complexity of $S(c)$.

6. In [5] the authors show that for the special cases of tree and square shapes, the minimum directed tile set problem is in P. For trees, it is straightforward to verify that the minimum tile set is always directed, so the answer is the same whether or not we restrict attention to directed tile sets. What is the complexity of the minimum tile set problem restricted to squares? The polynomial-time algorithm given in [5] crucially depends on the existence of a polynomial-time algorithm for the *directed shape verification* problem of determining whether a given tile system strictly self-assembles a given shape and is directed. Removing the directed constraint on this shape verification problem, even when restricted to the case of squares, makes the problem coNP-complete [6, 18, 24]. Perhaps this means that the minimum tile set problem restricted to squares is hard as well. On the other hand, since this problem is sparse,[13] Fortune's Theorem [16] implies that it cannot be coNP-hard (nor NP-hard by Mahaney's Theorem [27]) unless P = NP.

# References

[1] Scott Aaronson and John Watrous. Closed timelike curves make quantum and classical computing equivalent. *Proceedings of the Royal Society A: Mathematical, Physical and Engineering Science*, 465(2102):631, 2009.

[2] Zachary Abel, Nadia Benbernou, Mirela Damian, Erik D. Demaine, Martin Demaine, Robin Flatland, Scott Kominers, and Robert Schweller. Shape replication through self-assembly and RNase enzymes. In *SODA 2010: Proceedings of the Twenty-first Annual ACM-SIAM Symposium on Discrete Algorithms*, Austin, Texas, 2010. Society for Industrial and Applied Mathematics.

[3] Leonard M. Adleman. Towards a mathematical theory of self-assembly. Technical report, University of Southern California, 2000.

[4] Leonard M. Adleman, Qi Cheng, Ashish Goel, and Ming-Deh Huang. Running time and program size for self-assembled squares. In *STOC 2001: Proceedings of the thirty-third annual ACM Symposium on Theory of Computing*, pages 740–748, Hersonissos, Greece, 2001. ACM.

[5] Leonard M. Adleman, Qi Cheng, Ashish Goel, Ming-Deh A. Huang, David Kempe, Pablo Moisset de Espanés, and Paul W. K. Rothemund. Combinatorial optimization problems in self-assembly. In *STOC 2002: Proceedings of the Thirty-Fourth Annual ACM Symposium on Theory of Computing*, pages 23–32, 2002.

---

[13]More precisely, if one takes a bit of care in encoding the problem, then it can be assumed sparse. Assume that each square $S$ has its lower left-corner at the origin and is encoded by a list of its points in binary using precisely $\lfloor \log |S| \rfloor + 1$ bits for each point, and the bound $c$ on the number of tile types is also encoded in binary using precisely $\lfloor \log |S| \rfloor + 1$ bits. Then a sparse set that is $\equiv_m^P$-equivalent to the minimum tile set problem for squares is $\{ \langle S, c \rangle \in \text{MinTileSet} \mid S \text{ is an } n \times n \text{ square with lower-left corner } (0,0), \text{ and } 1 \leq c \leq n^2 \}$. In fact we can even require $c \leq O(\log n / \log \log n)$ by a result of [4], but the trivial upper bound of $n^2$ suffices to obtain sparseness.

[6] Gagan Aggarwal, Qi Cheng, Michael H. Goldwasser, Ming-Yang Kao, Pablo Moisset de Espanés, and Robert T. Schweller. Complexities for generalized models of self-assembly. *SIAM Journal on Computing*, 34:1493–1515, 2005. Preliminary version appeared in SODA 2004.

[7] Sanjeev Arora and Boaz Barak. *Computational Complexity: A Modern Approach*. Cambridge University Press, 2009.

[8] Robert D. Barish, Rebecca Schulman, Paul W. Rothemund, and Erik Winfree. An information-bearing seed for nucleating algorithmic self-assembly. *Proceedings of the National Academy of Sciences*, 106(15):6054–6059, March 2009.

[9] Florent Becker, Ivan Rapaport, and Eric Rémila. Self-assembling classes of shapes with a minimum number of tiles, and in optimal time. In *FSTTCS 2006: Foundations of Software Technology and Theoretical Computer Science*, pages 45–56, 2006.

[10] Harish Chandran, Nikhil Gopalkrishnan, and John H. Reif. The tile complexity of linear assemblies. In *ICALP 2009: 36th International Colloquium on Automata, Languages and Programming*, volume 5555, pages 235–253. Springer, 2009.

[11] Erik D. Demaine, Martin L. Demaine, Sándor P. Fekete, Mashhood Ishaque, Eynat Rafalin, Robert T. Schweller, and Diane L. Souvaine. Staged self-assembly: Nanomanufacture of arbitrary shapes with $O(1)$ glues. *Natural Computing*, 7(3):347–370, 2008. Preliminary version appeared in DNA 13.

[12] Erik D. Demaine, Matthew J. Patitz, Robert T. Schweller, and Scott M. Summers. Self-assembly of arbitrary shapes with RNA and DNA tiles. Technical Report 1004.4383, Computing Research Repository, 2010.

[13] David Doty. Randomized self-assembly for exact shapes. In *FOCS 2009: Proceedings of the Fiftieth IEEE Conference on Foundations of Computer Science*, pages 85–94. IEEE, 2009.

[14] David Doty, Lila Kari, and Benoît Masson. Negative interactions in irreversible self-assembly. In *16th International Meeting on DNA Computing (DNA 16), Hong Kong, China, June 14-17, 2010.*, 2010.

[15] David Doty, Matthew J. Patitz, and Scott M. Summers. Limitations of self-assembly at temperature 1. *Theoretical Computer Science*. to appear. Preliminary version appeared in DNA 15.

[16] Steven Fortune. A note on sparse complete sets. *SIAM Journal on Computing*, 8(3):431–433, 1979.

[17] Kenichi Fujibayashi, Rizal Hariadi, Sung Ha Park, Erik Winfree, and Satoshi Murata. Toward reliable algorithmic self-assembly of DNA riles: A fixed-width cellular automaton pattern. *Nano Letters*, 8(7):1791–1797, 2007.

[18] Michael R. Garey and David S. Johnson. *Computers and Intractability*. W. H. Freeman, New York, 1979.

[19] Ming-Yang Kao and Robert T. Schweller. Reducing tile complexity for self-assembly through temperature programming. In *SODA 2006: Proceedings of the 17th Annual ACM-SIAM Symposium on Discrete Algorithms*, pages 571–580, 2006.

[20] Ming-Yang Kao and Robert T. Schweller. Randomized self-assembly for approximate shapes. In Luca Aceto, Ivan Damgård, Leslie Ann Goldberg, Magnús M. Halldrsson, Anna Ingólfsdóttir, and Igor Walukiewicz, editors, *ICALP 2008: International Colloqium on Automata, Languages, and Programming*, volume 5125 of *Lecture Notes in Computer Science*, pages 370–384. Springer, 2008.

[21] Ryan J. Kershner, Luisa D. Bozano, Christine M. Micheel, Albert M. Hung, Ann R. Fornof, Jennifer N. Cha, Charles T. Rettner, Marco Bersani, Jane Frommer, Paul W. K. Rothemund, and Gregory M. Wallraff. Placement and orientation of individual DNA shapes on lithographically patterned surfaces. *Nature Nanotechnology*, 3:557–561, 2009.

[22] James I. Lathrop, Jack H. Lutz, Matthew J. Patitz, and Scott M. Summers. Computability and complexity in self-assembly. *Theory of Computing Systems*. to appear.

[23] James I. Lathrop, Jack H. Lutz, and Scott M. Summers. Strict self-assembly of discrete Sierpinski triangles. *Theoretical Computer Science*, 410:384–405, 2009.

[24] Harry R. Lewis and Christos H. Papadimitriou. *Elements of the Theory of Computation*. Prentice-Hall, Inc., Upper Saddle River, New Jersey, 1998.

[25] Dage Liu, Sung Ha Park, John H. Reif, and Thomas H. LaBean. DNA nanotubes self-assembled from triple-crossover tiles as templates for conductive nanowires. *PNAS: Proceedings of the National Academy of Sciences of the United States of America*, 101(3):717, 2004.

[26] Kyle Lund, Anthony T. Manzo, Nadine Dabby, Nicole Micholotti, Alexander Johnson-Buck, Jeanetter Nangreave, Steven Taylor, Renjun Pei, Milan N. Stojanovic, Nils G. Walter, Erik Winfree, and Hao Yan. Molecular robots guided by prescriptive landscapes. *Nature*, 465:206–210, 2010.

[27] Stephen R. Mahaney. Sparse complete sets for NP: Solution of a conjecture of Berman and Hartmanis. *JCSS: Journal of Computer and System Sciences*, 25, 1982. Preliminary version appeared in FOCS 1980.

[28] Hareem T. Maune, Si-Ping Han, Robert D. Barish, Marc Bockrath, William A. Goddard III, Paul W. K. Rothemund, and Erik Winfree. Self-assembly of carbon nanotubes into two-dimensional geometries using DNA origami templates. *Nature Nanotechnology*, 5:61–66, 2010.

[29] Ján Maňuch, Ladislav Stacho, and Christine Stoll. Step-assembly with a constant number of tile types. In *ISAAC 2009: Proceedings of the 20th International Symposium on Algorithms and Computation*, pages 954–963, Berlin, Heidelberg, 2009. Springer-Verlag.

[30] Ján Maňuch, Ladislav Stacho, and Christine Stoll. Two lower bounds for self-assemblies at temperature 1. *Journal of Computational Biology*, 17(6):841–852, 2010.

[31] Sung Ha Park, Constantin Pistol, Sang Jung Ahn, John H. Reif, Alvin R. Lebeck, Chris Dwyer, and Thomas H. LaBean. Finite-Size, Fully Addressable DNA Tile Lattices Formed by Hierarchical Assembly Procedures. *Angewandte Chemie-International Edition*, 45(40):6607–6607, 2006.

[32] Sung Ha Park, Hao Yan, John H. Reif, Thomas H. LaBean, and Gleb Finkelstein. Electronic Nanostructures Templated on Self-Assembled DNA Scaffolds. *Nanotechnology*, 15:S525–S527, 2004.

[33] Matthew J. Patitz. Simulation of self-assembly in the abstract tile assembly model with ISU TAS. In *FNANO 2009: 6th Annual Conference on Foundations of Nanoscience: Self-Assembled Architectures and Devices (Snowbird, Utah, USA, April 20-24 2009)*, pages 209–219. Sciencetechnica, 2009.

[34] Paul W. K. Rothemund. *Theory and Experiments in Algorithmic Self-Assembly*. PhD thesis, University of Southern California, December 2001.

[35] Paul W. K. Rothemund and Erik Winfree. The program-size complexity of self-assembled squares (extended abstract). In *STOC 2000: Proceedings of the Thirty-Second Annual ACM Symposium on Theory of Computing*, pages 459–468, 2000.

[36] Paul W.K. Rothemund, Nick Papadakis, and Erik Winfree. Algorithmic self-assembly of DNA Sierpinski triangles. *PLoS Biology*, 2(12):2041–2053, 2004.

[37] Marcus Schaefer and Christopher Umans. Completeness in the polynomial-time hierarchy: Part I: A compendium. *SIGACTN: SIGACT News (ACM Special Interest Group on Automata and Computability Theory)*, 33(3):32–49, September 2002.

[38] Nadrian C. Seeman. Nucleic-acid junctions and lattices. *Journal of Theoretical Biology*, 99:237–247, 1982.

[39] David Soloveichik and Erik Winfree. Complexity of self-assembled shapes. *SIAM Journal on Computing*, 36(6):1544–1569, 2007. Preliminary version appeared in DNA 10.

[40] Larry J. Stockmeyer. The polynomial-time hierarchy. *Theoretical Computer Science*, 3(1):1–22, October 1976.

[41] Scott M. Summers. Reducing tile complexity for the self-assembly of scaled shapes through temperature programming. Technical Report 0907.1307, Computing Research Repository, 2009.

[42] Hao Wang. Proving theorems by pattern recognition – II. *The Bell System Technical Journal*, XL(1):1–41, 1961.

[43] Hao Wang. Dominoes and the AEA case of the decision problem. In *Proceedings of the Symposium on Mathematical Theory of Automata (New York, 1962)*, pages 23–55. Polytechnic Press of Polytechnic Inst. of Brooklyn, Brooklyn, N.Y., 1963.

[44] Erik Winfree. *Algorithmic Self-Assembly of DNA*. PhD thesis, California Institute of Technology, June 1998.

[45] Celia Wrathall. Complete sets and the polynomial-time hierarchy. *Theoretical Computer Science*, 3(1):23–33, October 1976.

[46] Hao Yan, Sung Ha Park, Gleb Finkelstein, John H. Reif, and Thomas H. LaBean. DNA-templated self-assembly of protein arrays and highly conductive nanowires. *Science*, 301(5641):1882, 2003.